\documentclass{aa}
\usepackage{amsmath}
\usepackage{graphicx}
\usepackage{txfonts}
\usepackage{natbib}
\bibpunct{(}{)}{;}{a}{}{,} % to follow the A&A style
\newcommand{\teff}{\mbox{${T}_{\rm eff}$}}
\newcommand{\logg}{\mbox{${\log g}$}}

\newcommand{\msol}{\mbox{${\rm M}_{\odot}$}}
\newcommand{\msun}{\mbox{${\rm M}_{\odot}$}}

\newcommand{\aaa}{\mbox{$\alpha$}}
\newcommand{\sct}{\mbox{$\delta$~Scuti\ }}
\newcommand{\simgt}{\lower.5ex\hbox{$\; \buildrel > \over \sim \;$}}
\newcommand{\simlt}{\lower.5ex\hbox{$\; \buildrel < \over \sim \;$}}

\begin{document}
\title{Effect of convective outer layers  modeling on non-adiabatic seismic observables of $\delta$~Scuti stars}
\author{J.~Montalb\'an\inst{1}, M.-A. Dupret\inst{2}}
\offprints{J.~Montalb\'an}
\institute{
Institut d'Astrophysique et de G\'{e}ophysique de l'Universit\'{e} de Li\`{e}ge, Belgium \and
Observatoire de Paris, LESIA, 92195 Meudon, France 
}

\date{}
\titlerunning{Effect of convection modeling on non-adiabatic seismic observables of $\delta$~Scuti stars}
\authorrunning{J. Montalb\'an and M.-A. Dupret}
  \abstract
   {The identification of pulsation modes in \sct\ stars is mandatory to constrain the theoretical stellar
models. The non-adiabatic observables used in the photometric identification methods depend, however,
on  convection modeling in the external layers.}
   { To determine how the treatment of convection in the atmospheric and sub-atmospheric layers affects
the mode identification, and what information about the thermal structure of the external layers 
can be obtained  from amplitude ratios and phase lags in  Str\"omgren photometric bands.  
   }
   { We  derive non-adiabatic parameters for \sct\ stars by using, for the first time, stellar models 
with the same treatment of convection in the interior and in the atmosphere.
We compute classical non-gray mixing length models, and as well non-gray ``Full Spectrum of Turbulence'' models.
Furthermore,  we 
compute the photometric amplitudes and phases of pulsation by using the  colors and the limb-darkening coefficents 
 as derived from the same atmosphere models used in the stellar modeling.
   }
   {We show that the non-adiabatic phase-lag is mainly sensitive to the thermal gradients in the external 
layers, (and hence to the  treatment of convection), and that this
  sensitivity is also  clearly reflected in the multi-color photometric phase differences.
}
   {}

   \keywords{convection -- stars: atmospheres -- stars: oscillations -- stars: evolution -- stars: variables: $\delta$Scuti}

\maketitle

\section{Introduction}

Asteroseismology uses the frequency of stellar pulsations as a probe of the interior stellar 
structure, by comparing the observed frequencies with those determined from stellar models. 
In order to constrain the physical parameters of a star from its pulsation frequencies, it 
is necessary to identify the modes of pulsation, that is, to determine the spherical
harmonic degree $\ell$, the azimuthal order $m$, and the radial order $n$ for each mode. 
This is far from trivial in \sct stars, as their excited modes  do not fall in the asymptotic range
and, furthermore, rotation and possible coupling  destroy any pattern regularity. 
As a consequence, mode identification is difficult and other complementary observables  must 
be used for this purpose.

The  mode identification  photometric methods are 
based on the analysis of luminosity variations in different photometric passbands, and  on the
dependence of their amplitudes and phases  on the spherical harmonic degree, $\ell$.
These techniques are based on the theoretical study by \cite{dziem77b} of
the light variations generated by  non-radial oscillation. 
Afterwards, several improvements 
were  made by \cite{BalonaStobi79}, 
\cite{StamforWatson81}, \cite{Watson88}, \cite{Garrido90} and 
\cite{Heyn94}. The main shortcoming of the above mentioned  approaches is that 
the non-adiabatic amplitude and phase  of the effective temperature are treated
as free parameters.
\cite{Cugier94} updated the method by using  Dziembowski's non-adiabatic 
code to derive amplitude and phases  and applied it to $\beta$~Cep stars.
This last approach
did not take into account the non-adiabatic iegenfunctions in the
 atmosphere layers, and  
assumes  the Lagrangian  temperature variation to be equal to the local 
effective temperature variation at the photosphere.
\citet{Dupret02} showed, however, that the temperature perturbation in the photosphere is  
rather different from the variation of the local effective temperature, 
because of the significant  optical depth variation  produced by the oscillation.

Since the  non-adiabatic observables (multi-color magnitude variation) strongly depend  on the variation 
of the atmospheric thermal structure, their accurate determination  requires to solve 
the adequate  non-adiabatic equations  in the stellar atmosphere as well.
\citet{Dupret03} included an improved  treatment of the oscillations in
the non-grey atmosphere. The application of their model to  $\beta$~Cephei and SPB stars \citep{Dupret03}
an  improvement in  mode identification.

Stars in the Hertzprung-Russell Diagram (HRD) region where $\delta$~Scuti stars are located
should have, according to theoretical  stellar models,
an external structure with one or two convective regions due to  H and He ionization.
Since these regions are very shallow, they are as well very over-adiabatic.
As a consequence, the thermal structure of the outer  layers, and therefore
the  photometric amplitude ratios and phases differences in different bands, are  very sensitive
to the  details of  convection modeling,  such as  \cite{BalonaEvers99} showed in the first
non-adiabatic study of \sct\ stars using the Dziembowski's code.

The ``standard model" of convection in stellar evolution is the mixing length
theory  \citep[MLT][]{Bohm58}, where turbulence is described by a
relatively simple model. This contains essentially one adjustable parameter,
 $\alpha$,  relating the mixing length to the local pressure scale height.
Several recent papers have discussed the effect  on the
non-adiabatic observables of  varying $\alpha$ \citep{jagoda03,Moya04,jagoda05}.
However, in those papers, the convection treatment is changed only in the
regions computed with the corresponding stellar structure code, and not in 
the atmospheric layers\footnote{The exact optical depth, $\tau=2/3$ or $\tau=10$
depends on the particular stellar evolution code used}, where the multi-color magnitude 
variations due to oscillations are generated.

The analysis of the effect of convection modeling on stellar structure and
evolution is not always obvious, since the available atmosphere 
models are tipically computed for a given theory of convection and  for a fixed
value of  $\alpha$--MLT. 
Luckily, \citet{Heiter02} have published new atmosphere model grids based
on Kurucz's {\sc ATLAS9} code.
In these new grid (thereafter {\sc NEMO} grid), the atmosphere models have been computed 
for different treatments of  convection: MLT (with two different values
of the $\alpha$ parameter), 
\citet{CM91} (CM), and \citet{CGM} (CGM) treatments 
(also known as  FST -- Full Spectrum of Turbulence--  treatments).
These choices  are extensively motivated in Sect.~2 of \citet{Heiter02},
and the main characteristics and differences among these convection treatments
will be described in Sect.~\ref{sec:theo}.

By including the NEMO grids in our model computations, we are able to study in a
self-consistent way the effect of convection treatment on the  oscillation  
amplitudes and phases  in different passbands. 
The analysis presented here is consistent at two levels: first, the
stellar models (interior and atmosphere) used in the non-adiabatic computations
have been obtained by using the same (either FST or MLT) convection treatment; and second, 
the color transformations tables, as well as the limb-darkening law derivation,  
are based on the same atmosphere models as those we included in the stellar structure modeling.
In Sect.~\ref{sec:theo}, we present  the models used in this study: the interior equilibrium models (Sect.~\ref{sec:models}),
the atmosphere models and limb-darkening (Sect.~\ref{sec:atm}), the non-adiabatic oscillation code  (Sect.~\ref{sec:non-adi})
 and the color transformations giving the observable properties of \sct\ pulsations (Sect.~\ref{sec:colorintro}).
 In Sect.~\ref{sec:convectionA}, we  provide an outline of the problem of the modeling of convection in A-type stars.

 One of the main results of \citet{Heiter02} is that
MLT models with a small mixing length parameter (e.g., $\alpha \sim 0.5$)
are equivalent, in the atmospheric region where the observed flux
originates, to FST models. Both treatments predict, indeed, a low convective efficiency for these 
layers. The thermal structure in the subphotospheric layers for models with MLT($\alpha=0.5$) and FST may be,
however, quite different.  We will study in this paper if the non-adiabatic features of $\delta$-Scuti oscillations
are sensitive to these differences.

In Sect.~\ref{sec:results} we  compare the results obtained with MLT and FST treatments and explain the origin of the 
differences. We  analyze as well the respective weights of the  over-adiabatic structure
and of the color transformations  on the theoretical amplitude ratios and on phase differences in 
 Str\"omgren  photometric passbands.
Finally, we consider in Sect.~\ref{sec:ANlyn} the application to a particular \sct star: 
AN~Lyn, and, in Sect.~\ref{sec:conclu}, we present our conclusions.

\section{Theoretical models}
\label{sec:theo}

\subsection{Stellar models}
\label{sec:models}
The stellar models were computed by means of  the stellar evolution
code {\sc ATON3.0}, recently updated to be used in asteroseismology modeling \citep{Dantona05}. 
We  computed models for a metal mass fraction $Z$=0.02 and assumed a helium mass fraction $Y=0.28$. 
 At temperatures $T\geq$~12000~K we adopt the OPAL radiative opacities
($\overline\kappa$) \citep{Iglesias96}  for the solar Z--distribution
from \citet{Grevesse93}. In the  high--density ($\rho$) regions the opacities
are linearly extrapolated (\rm $\log~\overline\kappa\,\,vs. \log~\rho$), and
harmonically added to conductive opacities by \citet{Itoh93}. At
lower temperatures we use \citet{AF94} molecular opacities
(plus electron conduction in full ionization) for the same H/He ratios as in
the OPAL case.
A complete description of the equation of state (EOS) used in {\sc ATON3.0}   code is
given in \citet{Dantona05}. In short this code  uses  OPAL(2001) EoS \citep{Rogers02} %in short?? 
and \citet{Saumon95} EoS  for low temperature and high density domain.

Several treatments of convection are available in ATON3.0: the classic mixing length theory (MLT) model,
with the \citet{coxgiuli} formulation, and two different FST models, according to \citet{CM} and \citet{CGM}
approaches. A detailed description of the convection modeling in ATON3.0
is provided by \citet{ventura98}. For the present study, we adopt MLT and \citet[thereafter FST or CGM]{CGM} models.
Except for explicitly mentioned cases, the $\alpha$-MLT  and the $\alpha^*$-FST parameters are the same
as those in the atmosphere models which were used to derive  external boundary conditions.

As surface boundary conditions \citep[for detail, see][]{montalban04} we take the 
{\sc NEMO} atmosphere grids. As  mentioned above,  these new atmosphere 
models have been computed using MLT  model (with $\alpha=0.5$ and,  for a smaller 
range of parameters,  $\alpha=1.25$ as in classic ATLAS9 model), 
and, as well,  CGM with $\alpha^*=0.09$ \citep{CGM}. 
That allowed us to consistently build  stellar models (interiors and atmospheres).
The boundary is  located at optical depth $\tau_{\rm ph}=10$, as suggested by
\citet{Heiter02}), except  when MLT models are computed with a $\alpha$--MLT value in the 
interior different from  that of the  atmosphere. In this case,
the boundary is located at  $\tau_{\rm ph}=1$, to avoid  mixing of different 
convection efficiencies in the overadiabatic region.

Helium and metal gravitational settling  are not included in these computations. 
All the models are computed following the evolution from the pre-main sequence  (PMS) phase.
As  numerical resolution is concerned, the
number of models for the PMS phase is of the order of 600, and the evolution from the ZAMS to the TAMS
is  done in $\sim$350 models, besides,  2500--3000 mesh points are used to describe the stellar structure.

Before applying non-adiabatic computations, the stellar structure of the equilibrium
model is extended by matching the atmosphere structure  (from $\tau_{\rm ph}$ up to $\tau=10^{-4}$)  
as obtained by  interpolation in the grid  of atmosphere models.

\subsection{Convection modeling in A-type stars}
\label{sec:convectionA}

Standard models for A-type stars predict stellar structures with a convective core, and one or two 
external convective regions, the inner one  corresponding to  HeII ionization,  and the
shallower and outer one to that of HI and HeI. 
Convection in the central region is almost adiabatic, while
the external convective regions are  highly overadiabatic, but very thin and shallow. 
As a consequence, the details of convection in A-type stars (except for the
overshooting that changes the dimension of the convective core)  have a negligible effect on the 
their location in the HR diagram.

Nevertheless, the convective heat transport changes the temperature gradient, and  affects, therefore,
 the spectral line profiles and  the emission in  different photometric bands. 
In fact, there are spectroscopic evidences  of the presence of convection in A-type stars, such
as the peculiar behavior of Balmer lines as a function of temperature, the shape of bisectors, or
chromospheric activity \citep[see][for a review]{Smalley04}.
These observations suggest that convective velocity fields of several km/s are present in A-stars, 
while the temperature gradient must be close to the radiative one, except for the coolest 
objects ($\sim 7000$~K).

It has been  already shown by \citet{Fuhrmann93},  \citet{Fuhrmann94}, \citet{vantveer96},
\citet{vantveer98}, \citet{Barklem02}, \citet{SmalleyKupka97}, \citet{SmalleyKupka98},
\citet{Smalleyetal02},  
that spectroscopic and photometric features seem  better fitted by adopting low efficiency convective models
(for instance, MLT with \aaa=0.5 or CGM) which provide temperature gradient close to the radiative one.
Nevertheless, neither MLT or CGM succeed in matching all the photometric features in the A-F stars  effective temperature domain. 

 Other observational evidences in favor of low efficient convection in A-type stars has been
provided  by \cite{jagoda03} and \cite{jagoda05} in their fits of the seismic features of several
\sct\ stars.
 
Either, non-local models of convection \citep{KupkaMontgomery02} 
or 2D and 3D numerical simulations \citep{Ludwigetal99,Freytag04}
indicate that, in order to reproduce the convective flux in the external layers,  
the value of the  MLT--\aaa\ parameter should vary from 0.5  at 8000~K to $\sim$1 or larger at 7000~K.
On the other hand,  the \aaa\ value required to reproduce the flux in the HeII convective zone should also be  four times larger than that used in the HI, HeI convective zone.
Fortunately (or unfortunately) for what concerns the present study,  \citet{Moya04} has shown that 
a change of  convection efficiency  in the HeII 
convective region does not affect the excited modes of \sct stars.

In the computation presented in this paper we  use the Mixing Length Theory (B\"ohm--Vitense 1958), and a
 the ``Full Spectrum of Turbulence" formalism (CGM). 
 The main improvement of CGM with respect to  MLT is that while the latter treats the heat transport  
 by mimicking the spectral distribution of eddies by one ``average" eddy (reliable only for
high viscosity fluids), the former overcomes the one--eddy approximation by using a
turbulence model to compute the full spectrum of a turbulent convective flow.
\citet{Canuto96} showed that a consequence of this different heat transport is that
 in the limit of highly efficient convection ($S
\gg 1$, where $S$ is the convective efficiency, see e.g. CM for details)  MLT
underestimates the convective flux, while -- in the low efficiency limit -- MLT
overestimates it. So, CGM convective fluxes are $\sim 10$ times larger than the 
MLT ones for the
$ S\gg 1$ limit while, in the low efficiency limit ($S \ll 1$),  the CGM 
 flux is $\sim 0.3$ times the MLT one. This
behavior yields, in the over-adiabatic region at the top of a convection zone,
steeper temperature gradients for FST than for solar--tuned MLT models.
In inefficient convection, the  temperature gradient sticks to
the radiative one and starts splitting only when convection becomes
efficient.
Another difference between CGM and MLT concerns the length scale ($\Lambda$) for transport processes:
  $\Lambda_{\rm MLT}=\alpha \cdot H_{\rm p}$ and 
$\Lambda_{\rm CGM} = z + \alpha^* H_{\rm p, top}$, where $z$ is the distance to a boundary
of convective region, and the second term is a fine tuning parameter that 
allows  small adjustments, if exact stellar radii are needed ( e.g., in
helioseismology). As \citet{CGM} stress, the role  of $\alpha^*$
in solar FST models is radically different from that of  $\alpha$
in the MLT model. In fact, FST tuning affects only layers close to
the boundaries.  For inner layers, $z$ quickly grows and becomes much
larger than $\alpha^* H_{\rm p,top}$. In the case of A-type stars, however,
the small dimension of convective zones implies that this term could be 
a significant fraction of $z$.
For the models presented in this paper, we  use $\alpha^*=0.09$, as derived by \citet{CGM} from 
solar calibration \footnote{Note that this value was derived
for the solar calibration using grey-atmosphere as external boundary conditions} and adopted by
Heiter et al. (2002) for their  atmosphere models.

\subsection{Atmosphere models and limb-darkening}
\label{sec:atm}
 The original ATLAS9 code treats    convection by using MLT
with $\alpha=1.25$~  and thas the possibility of including a sort of overshooting
\citep[see][ for details]{Castelli97b,Castelli97a} 
The value $\alpha$=1.25 and the
option of ``approximate overshooting'' were originally adopted to fit the intensity spectrum
at the center of the solar disk and the solar irradiance. However,
\citet{Castelli97a} showed that these quantities are much more sensitive to the
overshooting--on mode than to the value of $\alpha$ itself  \citep[see also][]{Heiter02}.
The new grids of ATLAS9 atmospheres by \citet{Heiter02},  which we will use in our 
computations, introduce some  improvements with respect to the original ones  published by
\citet{Kurucz93} and \citet{Castelli97a},  in particular,
they provide  a finer grid spacing   ($\Delta$\teff, $\Delta \log g$), as well as a higher  vertical
resolution.
 For  the reader's  convenience, we recall the main characteristics of these grids:
\begin{itemize}
\item[i)] Three different models of convection are considered: MLT($\alpha=0.5$), FST according to CM and 
to CGM.
\item[ii)]  For FST models,  the vertical resolution in the atmospheric integration is increased
from 72 to 288 layers, spanning the range from $\log \tau_{\rm Ross} = -6.875$ to $\log
\tau_{\rm Ross} = 2.094$ (where $\tau_{\rm Ross}$ is the  average Rosseland optical depth).
\item[iii)] Effective temperature range: from  4000 to 10000~K, with $\Delta$\teff~=
200~K, and 
\item[iv)] gravity ( $\log g$) from  2.0 to 5.0,  with $\Delta \log g = 0.2$.
In the new ATLAS9-MLT grids, the convection is described as in \citet{Castelli96}  and
\citet{Castelli97a} (that is $V/A=\Lambda/6$ and $y=0.5$), but with
$\alpha=0.5$~ and without ``overshooting''.
\end{itemize}

In the CGM model atmospheres, the convective flux is computed as in \citet{CGM},
 but the characteristic scale length is defined as: $\Lambda =
\min(z_{\rm top} + \alpha^* H_{\rm p,top},  z_{\rm bot} + \alpha^* H_{\rm
p,bot} )$ where the index ``top" and ``bot" refers to top and bottom of the
convective region, and $\alpha^*=0.09$ (see footnote 2).

The photometric methods (Sec.~\ref{sec:colorintro}) used for $\delta$~Scuti  mode identification
are based on an analytical expression of the monochromatic 
oscillation amplitudes and phases. In this expression (Eq.~\ref{mag}) explicitly appear
the  weighted limb-darkening integrals 
($b_{\ell,\lambda}$, Eq.~\ref{weight}) and their derivatives  with respect to $\log$~\teff\ and \logg\ (Eqs.~\ref{dert} and \ref{derg}). 
The new and denser grid of model atmospheres allows to derived  smoother $b_{\ell,\lambda}$, and their derivatives as 
required for   mode identification method (Garrido 2000). 

In this paper we will use the 
limb-darkening coefficients (LDC) for the Str\"omgren photometric 
system, as derived by \citet{Barban03}  by using the quadratic-law, for a sample of new  atmosphere
models with effective temperature  between 6000 and 8500~K, \logg\ between 2.5 and 4.5, and solar
metallicity.  These authors determined, for each $uvby$ band, the best limb-darkening law
as well as the integrals $b_{\ell,\lambda}$ and  found similar results for
 CGM and MLT~(\aaa=0.5), but significant differences between CGM and the usual MLT (\aaa=1.25).
Besides, they  showed that
the differences in the LDC's follow those in the vertical temperature structure 
of the atmosphere model  due to the different treatments of  convective transport.
The differences between CGM and  MLT~(\aaa=1.25) models are
higher for low temperatures, but  the differences propagate toward high temperatures for model
atmosphere with higher gravities.
The differences between LDC's computed for MLT(\aaa=0.5) and CGM atmosphere models
are much smaller than for the MLT(\aaa=1.25), but they can reach
the same order of magnitude as for MLT(\aaa=1.25) when  intermediate gravity and low temperature ($\sim 7600$~K)
are considered.
As a general statement, the differences in the $b_{\ell}$ coefficients and derivatives computed with CGM and 
MLT~(\aaa=0.5) are much smaller ($\sim 5$\%)  than differences 
 obtained between computations with CGM and MLT~(\aaa=1.25) ($\sim 20$\%).

As for the LDC's the effect of convection treatment on $b_{\ell}$ is maximum at low effective temperature and low 
gravity models, and these large differences extend towards higher temperatures for models with high gravity.
Furthermore, the magnitude of the differences in $b_{\ell}$ due to the convection treatment is $\ell$--dependent.

The use of a higher resolution in optical depth in the model atmospheres and the different treatment of the 
convection 
imply significant changes in the 
LDC values. In this paper we
use these results to analyze the effect of LDC's and of the external thermal stratification on the theoretically
determined amplitude ratios and phases.

\subsection{Non-adiabatic treatment}
\label{sec:non-adi}

We do a non-adiabatic seismic analysis of our models by using the code MAD \citep{Dupret03}.
In this code the adiabatic iegenfunctions are used as trial input for the non-adiabatic computations.
The input model is a stellar model extended with the atmosphere corresponding to its \teff, \logg\ and 
adopted convection treatment. In this context, we use for the first time the \citet{Heiter02} atmosphere models.

Since the  convection-pulsation interaction implemented in MAD cannot be applied to the FST formalism, we adopt
in our study the frozen convection flux approximation, that is, we neglect the Lagrangian variation of the
convective luminosity (radial component) and the Lagrangian variation of the transversal component of 
the convective flux. This approximation is not appropriate to describe the red edge of $\delta$~Scuti 
instability strip. Nevertheless, given the comparative character of this work and the low 
efficiency of convection  of most of the models considered here \footnote{FST models and
MLT ones with \aaa\ parameter smaller than one.},  we expect  the convection-pulsation 
interaction to be not relevant  to this  study. 
The transfer equation treatment in the MAD code is different in the interior and in the 
pulsating atmosphere \citep{Dupret03}. In the latter,
the hypothesis at the base of the procedure is not justified  for convective regions, and therefore, the 
match layer between the interior and the atmosphere  must be located outside the convective envelope.
Even if in the extended stellar model the match between atmosphere and interior is done in a convective region,
and the properties of a given  layer  come formally from the atmosphere model, the non-adiabatic treatment
reserved to the atmosphere will be apply only to layers such that the convective to radiative luminosity ratio is
smaller than $10^{-10}$.

These computations make possible to derive the phase lag ($\psi_{\rm T}$) 
 between the local relative variation of the effective temperature and the relative radial displacement 
($\delta\,r$) and the corresponding  amplitude ratio
($f_{\rm T}$),
which are basic ingredients to determine the magnitude variations in different photometric passbands (see
next section).

\begin{figure}
\resizebox{\hsize}{!}{\includegraphics{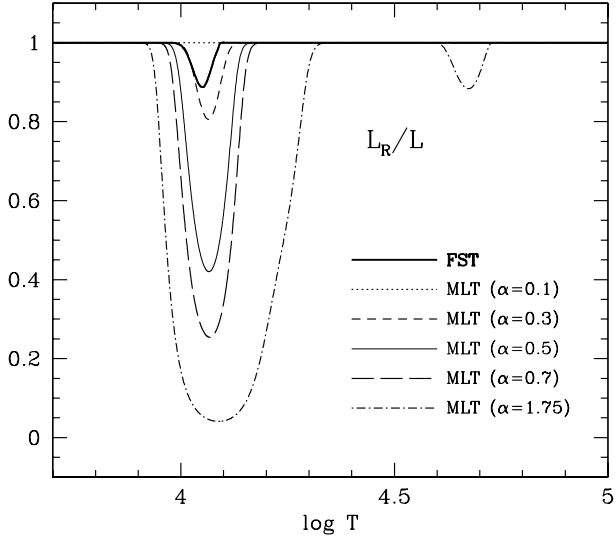}}
\caption{Fraction of luminosity transported by radiation with respect to total luminosity as 
a function of temperature for several 1.8~\msun\ models in the same HRD location (\teff=7400~K, \logg=3.98) 
but computed with different treatments of convection. The FST model has been computed with FST convection in 
the interior and FST atmospheres by \citet{Heiter02}. The MLT($\alpha=0.5$) model implies 
\aaa=0.5 in the interior and  the corresponding \aaa=0.5 \citet{Heiter02} atmosphere.
The MLT models with $\alpha\neq 0.5$ were computed with the indicated \aaa\ value in the interior
up to the optical depth $\tau=1$ and MLT($\alpha=0.5$) atmosphere model for $\tau < 1.$}
\label{lrl}
\end{figure}

\begin{figure}
\resizebox{\hsize}{!}{\includegraphics{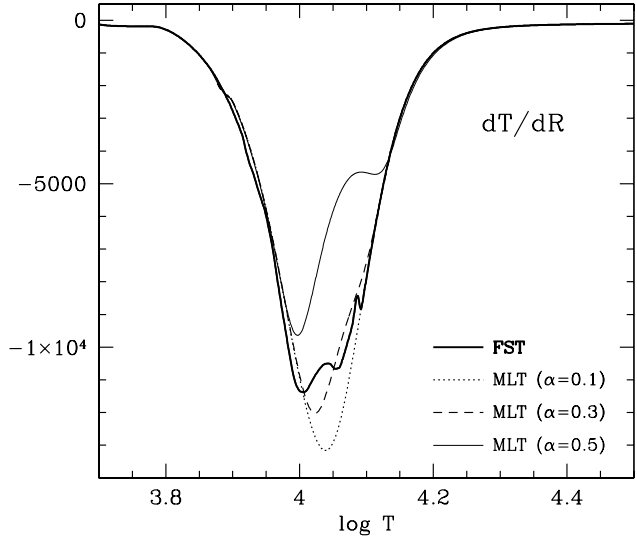}}
\caption{Temperature gradient as a function of temperature inside the star, for some of the
models in Fig.~\ref{lrl}.}
\label{grad}
\end{figure}

\subsection{Multi-color photometry}
\label{sec:colorintro}

 The non-adiabatic quantities 
$\psi_{\rm T}=\phi(\delta T_{\rm eff}/T_{\rm eff})-\phi(\delta r/R)$, $f_{\rm T}=|\delta T_{\rm eff}/T_{\rm eff}|/|\delta r/R|$ 
and $f_{\rm g}=|\delta g_{\rm eff}/g_{\rm eff}|$ 
(where $g_{\rm eff}$ is the effective  gravity and $\delta\,g$ the 
variation of gravity at the photosphere) can be related to 
 photometric observables such as the  amplitude ratios and phase differences between magnitude variations 
in different photometric passbands.  It has been shown, in fact, \citep[e.g.][and references therein]{Watson88}
that, in a  one-layer linear approximation, the magnitude variation at a wavelength $\lambda$ produced by
a stellar pulsation with spherical harmonic degree $\ell$,  azimuthal order $m$ and   
angular frequency $\sigma$ is given by
\begin{eqnarray}
\delta m_{\lambda} & = & -\frac{2.5}{\ln 10}\,a\,P_{\ell}^m (\cos i)\,b_{\ell \lambda}\,
\big\{ -(\ell-1)\,(\ell+2)\,\cos \,(\sigma t) \nonumber \\
& + & f_{\rm T}\,\cos\,(\sigma t + \Psi_{\rm T})\,(\alpha_{\rm T \lambda}+\beta_{\rm T \lambda}) \nonumber  \\
& - & f_{\rm g}\, \cos\,(\sigma t)\,(\alpha_{\rm g \lambda}+\beta_{\rm g \lambda})
\big\}
\label{mag}
\end{eqnarray}

\noindent
where $i$ is the inclination angle between the stellar rotation axis and the observer 
line of sight; $a$ is the amplitude  of the relative radial displacement at the photosphere and 
$P_{\ell}^{m}$ is the associated Legendre function of degree $\ell$ and 
azimuthal order $m$. $b_{\ell \lambda}$ is related to  the limb-darkening law by

\noindent 
\begin{equation}
b_{\ell \lambda}=\int_0^1\,h_{\lambda}\,\mu\,P_{\ell}\,d\mu
\label{weight}
\end{equation}

\noindent where $h_{\lambda}$ is the normalized limb-darkening function, and $\mu =\cos \theta$ with $\theta$
the angle between the line of sight and the normal to the local stellar surface. 
The rest of quantities appearing in Eq.~\ref{mag}: $\alpha_{\rm T \lambda}$, $\beta_{\rm T \lambda}$,
$\alpha_{\rm g \lambda}$ and $\beta_{\rm g \lambda}$ are derived from the appropriate atmosphere
model with effective temperature \teff\, gravity  \logg\ and given convection treatment:

\begin{equation}
\alpha_{\rm T \lambda}=\frac{\partial \ln F^+_{\lambda}}{\partial \ln T_{\rm eff}}; \,\,\,\,
\beta_{\rm T \lambda}=\frac{\partial \ln b_{\ell \lambda}}{\partial \ln T_{\rm eff}}; 
\label{dert}
\end{equation}

\begin{equation}
\alpha_{\rm g \lambda}=\frac{\partial \ln F^+_{\lambda}}{\partial \ln g_{rm eff}}; \,\,\,\,
\beta_{\rm g \lambda}=\frac{\partial \ln b_{\ell \lambda}}{\partial \ln g_{rm eff}}.
\label{derg}
\end{equation}

An appropriate way to test multi-color theoretical predictions is to construct phase-amplitude 
diagrams corresponding to well-chosen combinations of photometric bands \citep[e.g.][]{Garrido90}.
 The theoretical results  corresponding to modes of different spherical degrees $\ell$ occupy well 
distinct regions. 
This makes possible the   identification of $\ell$ by searching for the best fit between theory and 
observations. The  non-adiabatic quantities $f_{\rm T}$ and $\psi_{\rm T}$, which 
play a mayor role in Eq.~\ref{mag}, are highly sensitive to the characteristic of the convective 
envelope (see. Fig.~\ref{ftte} and \ref{psitte}). 
On the other hand, \citet{Barban03} showed  that the sensitivity of $b_{\ell \lambda}$ and its derivatives
to the convection model is different for different  photometric passbands and different degree $\ell$.
We expect, therefore, that  constraints on the characteristics of the convective 
envelope could be obtained from very precise multi-color photometric observations.

\section{General results for \sct stars}

\label{sec:results}
\subsection{Equilibrium models}

As  already mentioned,  thanks to the shallow convection envelope in A-type, the 
HRD location of a stellar model of fixed mass and chemical composition is not affected by the
treatment of non-adiabatic convection layers. This fact allows us to easily separate the effect of 
convection. We have computed models for 1.8 and 2.0~\msun\ with FST formalism, and with MLT($\alpha=0.5$).
We have also computed MLT models with $\alpha_{\rm int}=$0.1, 0.3, 0.7 and
1.75\footnote{ We would like to call attention on the fact that since the available atmosphere models
use MLT($\alpha$=0.5)  or CGM,  we are able to compute only ``complete'' MLT(\aaa=0.5) 
 and ``complete'' FST models (the same treatment of 
convection in the interior structure and in the atmosphere).
The internal structure of  models labeled with MLT~~$\alpha$=0.1, 0.3, 0.7 and 1.75 
has been  computed until $\tau=1$ with these $\alpha$ values, and with 
boundary conditions for temperature and pressure ($(T,P)(\tau=1)$) provided by the 
the MLT($\alpha$=0.5) atmosphere model.}. 
Even if for a given central hydrogen content, all the computations yield the same point in the HRD, the
characteristic of the external layers are affected  by different convection 
treatments. The two main physical quantities affected by this treatment are the energy fraction  transported by
convection (Fig.~\ref{lrl}) and the temperature gradient (Fig.~\ref{grad}).
As it is well known, the energy  fraction 
transported by convection increases as  $\alpha$ increases, while the  temperature gradient  decreases.
On the other hand, as pointed out by \citet{Canuto96},   the energy fraction 
transported by convection in  FST models is much smaller than that of MLT treatment in the case of low efficiency convection.

\begin{figure}
\resizebox{\hsize}{!}{\includegraphics{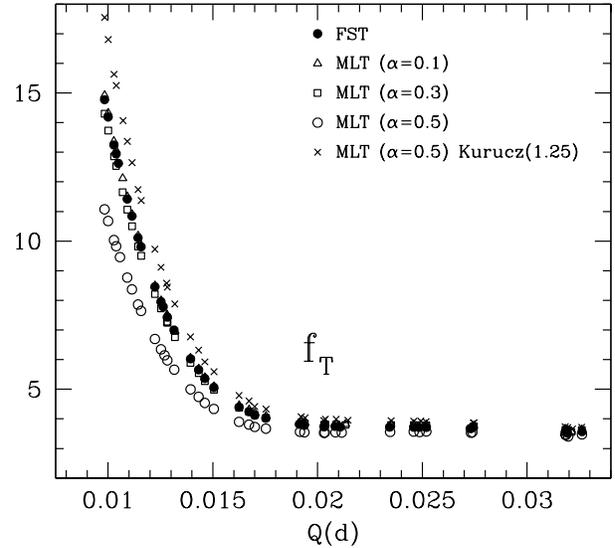}}
\caption{The non--adiabatic quantity $f_{\rm T}=|\delta T_{\rm eff}/T_{\rm eff}|$ as  a function
of the pulsation constant $Q$ (in days) for modes with spherical degree $\ell=0,1,2$ and 3.
These modes correspond to 1.8~\msun\ models in the same HRD location 
and computed with different convection treatment.
The meaning of labels is the same as in Fig.~\ref{lrl}. The label MLT (\aaa=0.5) Kurucz (1.25)
refers to a model computed with MLT(\aaa=0.5) in the interior computation and
\citet{Kurucz93}'s atmosphere models.}
\label{ft}
\end{figure}

\begin{figure}
\resizebox{\hsize}{!}{\includegraphics{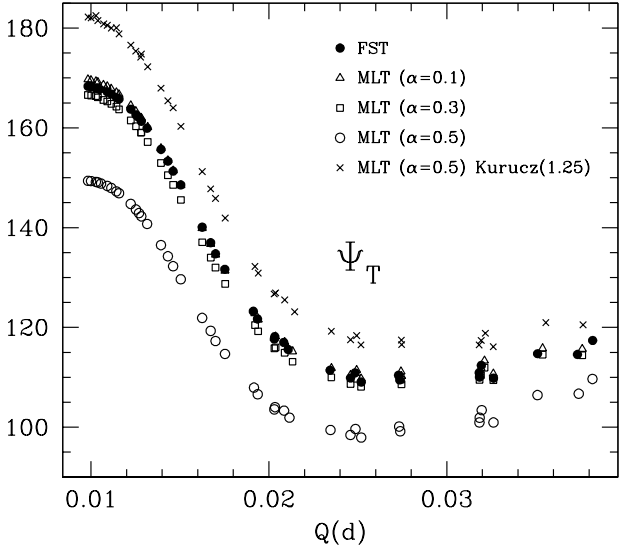}}
\caption{As in Fig.~\ref{ft} for the phase-lag (in degrees) between the local effective temperature
variation and  the radial displacement.}
\label{psit}
\end{figure}

In next section we analyze the consequences of these different structures on the non-adiabatic observables.

\subsection{Non-adiabatic results}
\label{nar}
We present now the non-adiabatic results obtained for models with a given mass ( 1.8~\msol) and effective temperature
$\sim$7400~K. 
In Figs.~\ref{ft} and \ref{psit}, we plot the relative amplitudes ($f_{\rm T}$) and phases ($\psi_{\rm T}$)
as a function of the  pulsation constant $Q$. 
$Q=P \: \sqrt{(R_{\sun}/R)^3(M/M_{\sun})}$, with $P$ being the period in days.
In \sct\ stars $Q \simeq 0.033$~days for the fundamental radial mode. 
The models used in  Figs.~\ref{ft} and \ref{psit} are the same as in Figs.~\ref{lrl} and \ref{grad} for
the same kind of convection, for all of them $\tau_{\rm dyn}/\tau_{\rm dyn,\sun}=\sqrt{(R/R_{\sun})^3(M_{\sun}/M)}=3.104282$.

The results presented in Figs.~\ref{ft} and \ref{psit} confirm that the non-adiabatic quantities
 (particularly the phase-lag $\psi_{\rm T}$) are very sensitive to the
convection treatment. As the value of the MLT-$\alpha$ parameter
increases, the phase-lags decreases. Furthermore,  for low values of the MLT-$\alpha$ parameter, the 
$\psi_{\rm T}$ value  for  high radial order modes 
remain close to the adiabatic one ($180\degr$), as already discussed by  \citet{Moya04} and \citet{Dupret05}. 

$f_{\rm T}$ and $\psi_{\rm T}$ values for FST model  are in between those obtained for  
 MLT($\alpha=0.1$) and MLT($\alpha=0.3$) models. 
In all our models except one, we use  NEMO atmosphere models with a
low efficient convection (FST or MLT--$\alpha$=0.5)).
The exception is for the model with an interior structure
computed with MLT($\alpha=0.5$) up to $\tau=1$, and 
with the  boundary conditions and the atmosphere
structure  given by $\alpha=1.25$ Kurucz's atmosphere models. 
The results provided by this last model and by the ``complete'' MLT($\alpha=0.5$) model
are significantly different.  This shows the relevance of 
 treating consistently the interior and the atmosphere.

We remark  that the pulsation frequencies are
almost identical for the  different  convection treatments. The reason is
that these models only differ in the outermost layers 
which have no weight at all on the frequencies of low order p-g modes.
As the effect of temperature gradient in mode excitation is concerned,  we state
that for models with very inefficient convection, i.e. FST, as well as MLT(\aaa=0.1) and MLT(\aaa=0.3),
only modes with $Q>0.014$~d are excited, whereas for MLT(\aaa=0.5) this limit
moves to higher frequencies, and modes with  $Q > 0.0125$~d  are excited as well.

\begin{figure}
\resizebox{\hsize}{!}{\includegraphics{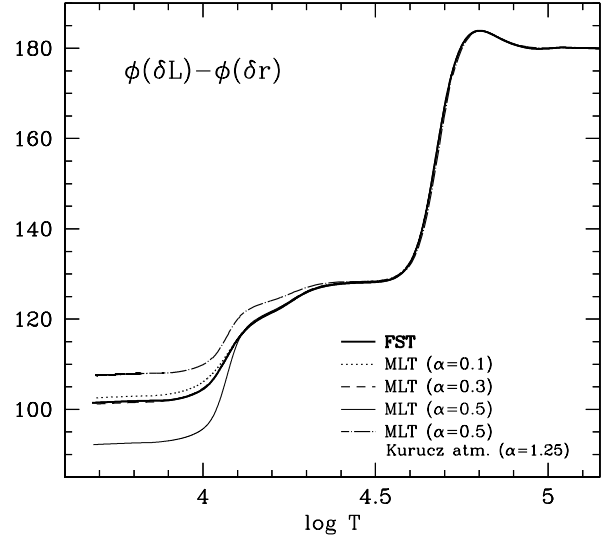}}
\caption{Luminosity phase-lag ($\phi(\delta L/L)-\phi(\delta r/r$) (in degrees) 
as a function of temperature inside the star for the radial $p_1$ mode corresponding to 
the same 1.8~\msun\ models than in previous figures.}
\label{dLfig}
\end{figure}

In order to understand the behavior of $\psi_{\rm T}$ for different convection models,
we plot in Fig.~\ref{dLfig} the phase-lag between the total luminosity perturbation
and the radial displacement ($\phi(\delta L)-\phi(\delta r)$) for the radial fundamental mode
 in  the same models as in Figs.~\ref{ft} and \ref{psit}. 
In deep regions ($\log T >5$) the pulsation is quasi-adiabatic and the phase-lag is $180\degr$.
A first phase-lag, quasi-independent of the convection treatment,
occurs at $\log T = 4.8$--4.6,  in the He\,{\sc ii} partial ionization zone, and
a second one appears in the H partial ionization zone ($4.2\ge\log T\ge 4.$).
This second  phase-lag  increases as the MLT $\alpha$ parameter increases, which leads to
phase-lag values close to  $0\degr$ at the stellar  surface. 

As shown in \cite{Dupret05}, the phase-lags can be interpreted on the basis of the following integral expression:

\begin{eqnarray}
\Delta\phi(\delta L) &\cong& - \int\Re\left\{\frac{\delta s /c_{\rm v}}{\delta L/L}\right\}
\frac{c_{\rm v}T\sigma}{L}{\rm d}m  \nonumber \\
&=& - \int\Re\left\{\frac{\delta s /c_{\rm v}}{\delta L/L}\right\}
\frac{4\pi r^2\rho c_{\rm v}T\sigma}{L\,{\rm d}\ln T/{\rm d}r}\;{\rm d}\ln T\,.
\label{phaselageq}
\end{eqnarray}

This equation makes evident the different contributions to the phase-lag through the star.
Thus, a significant contribution can be generated in a region where the thermal-relaxation time
is of the same order or larger than the pulsation period, and/or in a region where the pulsation is
highly non-adiabatic. This last situation is found in the convective zone coinciding with the H 
partial ionization zone (HCZ), where $\Re\{\delta s\:L / (c_{\rm v}\delta L)\}$ has a high value
due to the  the large opacity bump and to the temperature gradient.
The sensitivity of the phase-lag to the convection treatment can be made more obvious by noting
that the iegenfunctions $\delta L/L$ when written as a function of $T$ instead of as a function of mass, 
are essentially the same for different convection  treatments in the equilibrium models.
The main  reason is that the opacity is essentially a function of temperature and is not sensitive 
to $\alpha$ for a given interval or temperature \citep[see also][]{Moya03,Dupret05}.
As shown in Fig. \ref{grad}, what  significantly 
changes from one model to another is the temperature gradient.  When $\alpha$ increases, 
the temperature gradient in the HCZ decreases quickly.
According to Eq.~\ref{phaselageq} (second line) 
the phase-lag contribution from this zone  increases, therefore, with $\alpha$, and hence, 
the phase-lag can be considered as an indicator of the temperature gradient in the HCZ.

\begin{figure}
\resizebox{\hsize}{!}{\includegraphics{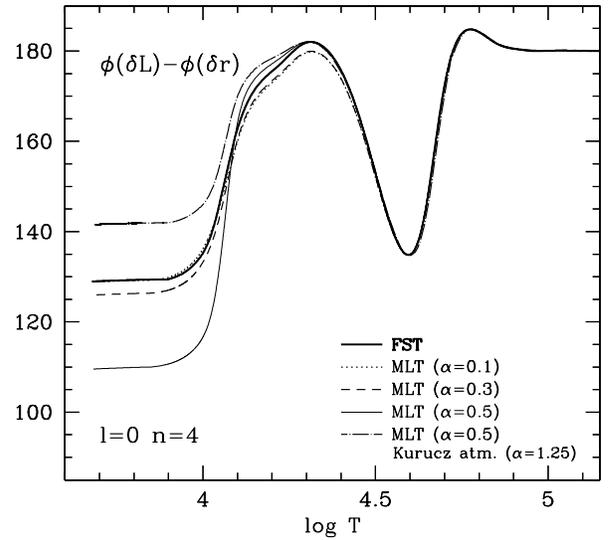}}
\caption{As Fig.~\ref{dLfig} for the n=4 overtone.}
\label{dLfigp4}
\end{figure}

In Fig. \ref{grad}, we see that
the mean temperature gradient of the FST model is between the temperature gradients of the MLT
models with $\alpha=0.1$ and $\alpha=0.3$, and consequently the phase-lags predicted by FST models 
are between the $\alpha=0.1$ and $\alpha=0.3$ MLT ones (see Figs~\ref{dLfig} and ~\ref{dLfigp4}).

It is well known that  different modes probe the different layers of stellar interiors. 
Similarly we can expect, in principle, that the non-adiabatic phase-lags predicted for different modes
probe the temperature gradient at different depths. It is evident from Fig. \ref{grad} that the exact shape of the
temperature gradient predicted by FST models cannot be reproduced by MLT models, whatever the value
of $\alpha$. Consequently, the phase-lag predicted by FST models cannot be reproduced by the same
MLT model in the whole  frequency spectrum. 
 This can be seen in Fig.~\ref{psit}. For periods close to that of the  fundamental radial mode,
  the FST phase-lags are close to the MLT($\alpha=0.3$)  ones. But for smaller 
periods, the FST phase-lags are closer to the $\alpha=0.1$ MLT ones. In Fig.~\ref{dLfigp4} we plot
$\phi(\delta L)-\phi(\delta r)$ of
the  p$_4$ modes computed for the same models of   
Fig.~\ref{dLfig}. As in the case of the p$_1$ mode, we have ${\rm d}\phi(\delta L)/{\rm d}r<0$ in the partial ionization zones of He and H, but, at variance with it,  ${\rm d}\phi(\delta L)/{\rm d}r>0$ in the
intermediate region between the He and H partial ionization zones, so that the phase-lag of the  p$_4$ mode
is closer to $180\degr$. These positive values of ${\rm d}\phi(\delta L)/{\rm d}r$ are associated with the damping 
of the modes in this region. Different shapes of the iegenfunctions, depending on the mode, imply a different 
sensitivity of the phase-lag to the convection treatment.  Nevertheless, the differences between
the phase-lags of the FST and the $\alpha=0.1$, $\alpha=0.3$ MLT models are small whatever the mode,
and it would be too optimistic to conclude that the present precision of observations for the phase-lags would 
make possible to discriminate between FST and MLT models with the lowest $\alpha$ values 
(see  discussion below).

\begin{figure}
\resizebox{\hsize}{!}{\includegraphics{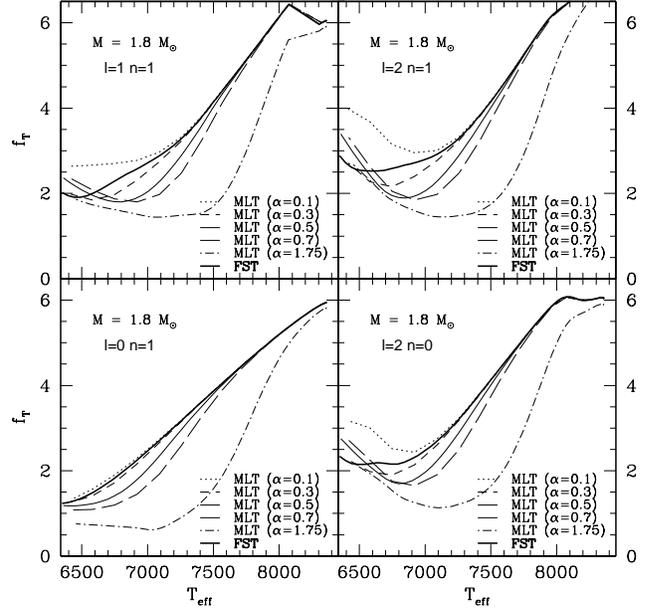}}
\caption{The non-adiabatic quantity $f_{\rm T}=|\delta T_{\rm eff}/T_{\rm eff}|$  for four different
modes ($\ell=0$, $n=1$; $\ell=1, n=1$; $\ell=2$, $n=0$ and $n=1$) 
as a function of the effective temperature for the MS evolution of 1.8~\msun\ models computed with different
convection treatments. The meaning of the labels is the same as in Fig.~\ref{lrl}. }
\label{ftte}
\end{figure}

\begin{figure}
\resizebox{\hsize}{!}{\includegraphics{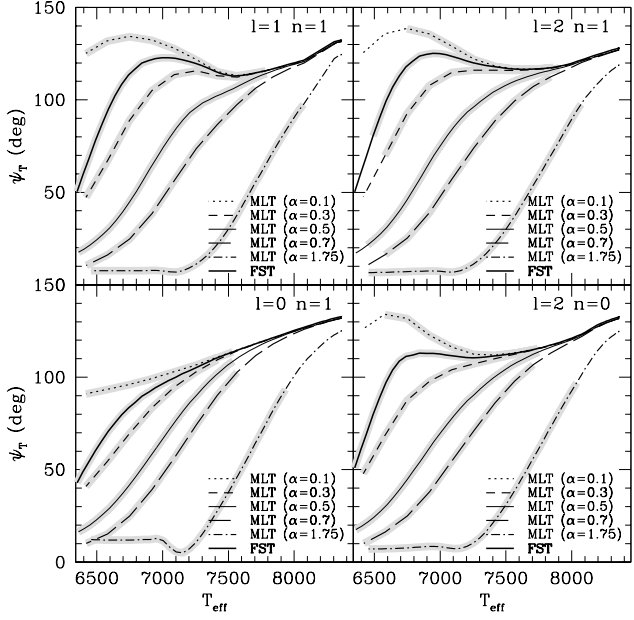}}
\caption{As Fig.\ref{ftte} for the temperature phase-lag: $\psi_{\rm T}$.  The gray bands indicate
the models for which the corresponding $\ell,n$ mode is excited.}
\label{psitte}
\end{figure}

\begin{figure}
\resizebox{\hsize}{!}{\includegraphics{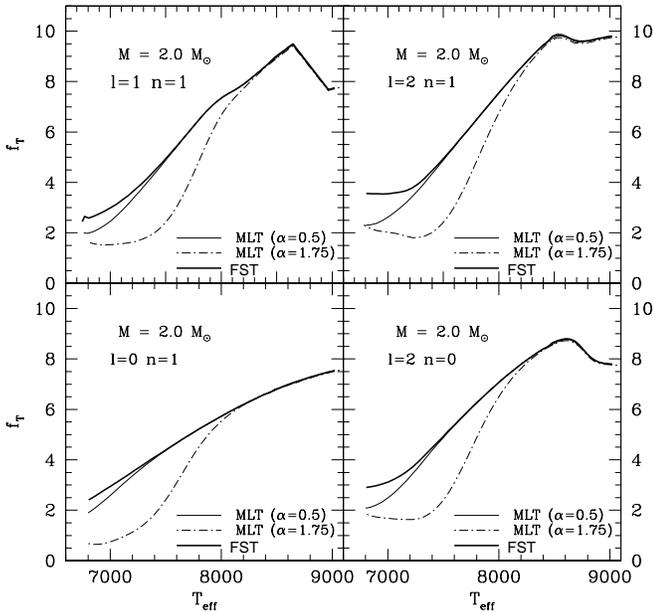}}
\caption{As Fig.~\ref{ftte} for 2.0~\msun\ models and for only three different convection
treatments.}
\label{ftte2}
\end{figure}

\begin{figure}
\resizebox{\hsize}{!}{\includegraphics{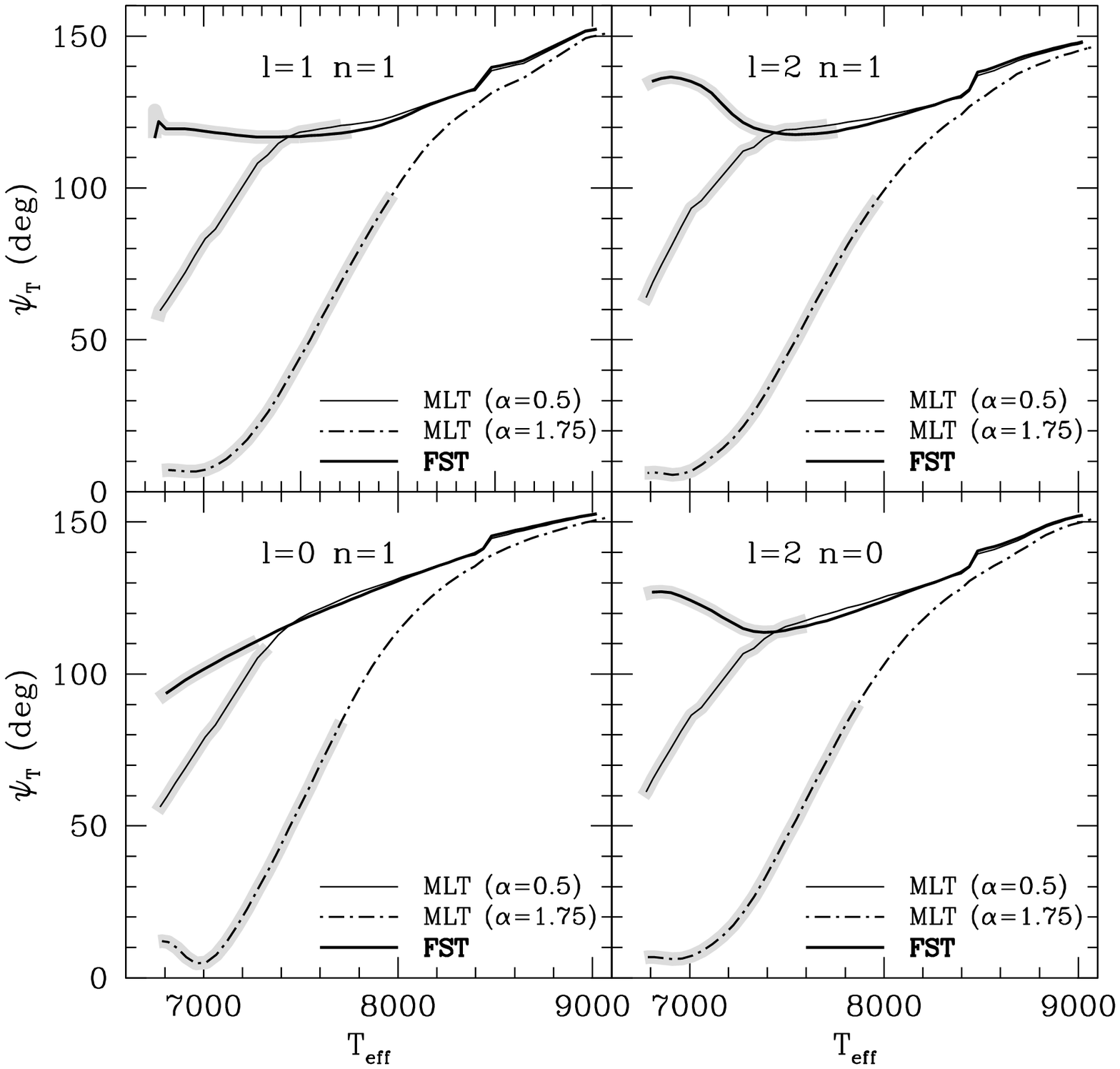}}
\caption{As Fig.~\ref{ftte2} for  the temperature phase-lag: $\psi_{\rm T}$.  The gray 
bands indicate the models for which the corresponding $\ell,n$ mode is excited.}
\label{psitte2}
\end{figure}

Finally, in Figs.\ref{ftte}--\ref{psitte2} we present the sensitivity of $f_{\rm T}$ and $\psi_{\rm T}$
to the convection treatment as a function of the effective temperature, for 1.8 and 2.0~\msun\ main-sequence
models. 
Comparing the MLT and FST curves, we see that the FST results are always between the $\alpha=0.3$ and $\alpha=0.1$
MLT results, closer to $\alpha=0.1$ for hotter models and to $\alpha=0.3$ for cold models.
It is also evident that $\psi_{\rm T}$ is much more sensitive than $f_{\rm T}$ to the convection 
treatment in the equilibrium model. Furthermore, the phase-lags decrease  as effective temperature
decrease, and tend to zero  more rapidly  as convection efficiency increases. 
The latter result agrees with those obtained by  \citet{Moya04} and 
by \citet{Dupret05} in the frozen convection approximation.
We remark that $\psi_{\rm T}$ differences among different convection treatments are larger for the 
cold models than for the hottest ones. 
These results can be understood as follows: when the effective temperature decreases, the size
of the convective envelope increases quickly and the gradient of temperature decreases. This is similar
to increasing $\alpha$ and, for the same reasons, the phase change between bottom and top of convection zone
increases. A second effect of decreasing  effective temperature is to increase the opacity bump. Consequently,
the associated bump of the non-adiabatic iegenfunctions is larger and  implies as  well larger phase changes. 

$\psi_{\rm T}$  for FST models follows the MLT($\alpha$=0.1) behavior until \teff$\sim 6800$~K,  then it changes 
suddenly its slope to follow the MLT($\alpha$=0.3)~~$\psi_{\rm T}$ curve.
This is explained by the properties of FST convection \citep[see e.g.][]{DantonaBV} which induce 
a rapid  change of the depth of the convective zone at that \teff.
That change from shallow to deep convection has the same effect on $\psi_{\rm T}$ as
the increasing of $\alpha$ parameter in the MLT models.
We note that  FST~~$\psi_{\rm T}$ values for \teff$\ge 6500$~K  models are between 90 and 140$\degr$, in 
good agreement with the typical observational values in \sct\ stars, and with the theoretical predictions
of time-dependent-convection models by \citet{Dupret05}.

For 2.0~\msun\ models, where the convective zone is shallower than in 1.8~\msun\ ones,
the effect of considering FST or MLT(\aaa=0.5) is smaller than for 1.8~\msun\ models, 
for both the amplitude ratio and  the phase-lag. Nevertheless, FST models show always 
$\psi_{\rm T} > 100\degr$ while for  MLT(\aaa=0.5) $\psi_{\rm T}$ decreases down to
$\sim 50\degr$ for lower effective temperatures.
High $\alpha$ values provide as before, too low  phase-lags.

\subsection{Multi-color photometry}
\label{sec:coloresult}

\begin{figure}
\resizebox{\hsize}{!}{\includegraphics{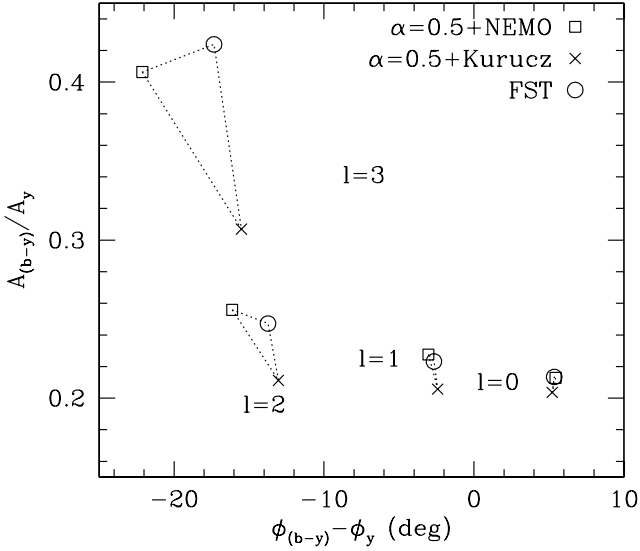}}
\caption{Color to light amplitude ratio {\it vs} the phase difference diagram for three 
1.8~\msun\ models: complete FST model (circles), complete MLT(\aaa=0.5) model(squares),
and  MLT(\aaa=0.5) model with \citet{Kurucz93}'s atmospheres   (crosses).
Different groups correspond to different $\ell$: $\ell=3$, upper-left;  $\ell=2$, lower-left;
 $\ell=1$, lower-middle and $\ell=0$, lower-right.}
\label{phi-a}
\end{figure}
\begin{figure}
\resizebox{\hsize}{!}{\includegraphics{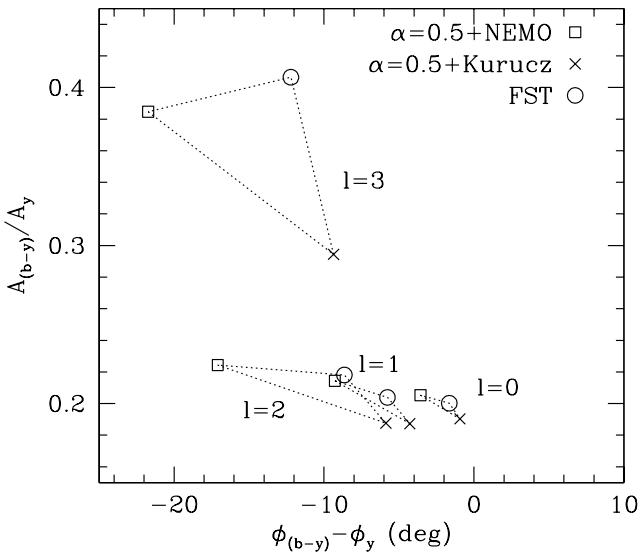}}
\caption{As Fig.~\ref{phi-a} but for a mode with  pulsation constant Q=0.015}
\label{phi-15}
\end{figure}
\begin{figure}[t]
\resizebox{\hsize}{!}{\includegraphics{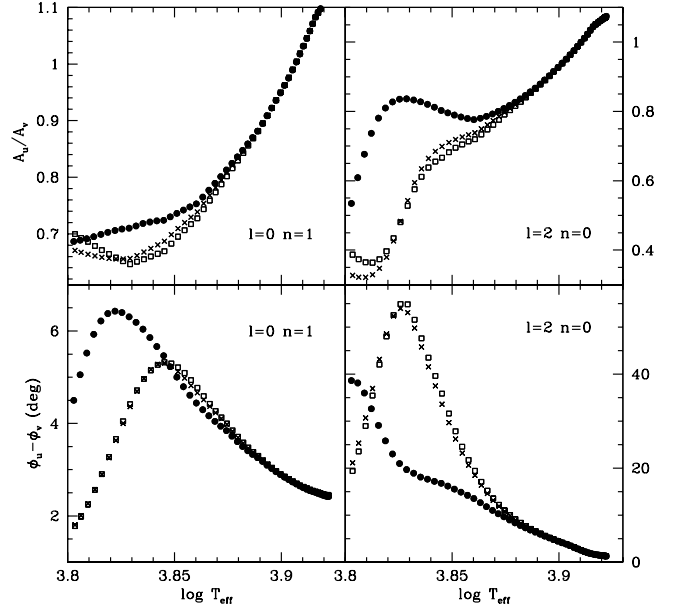}}
\caption{Photometric amplitude ratio (upper panels) and phase-lag (lower panels)
  for the $u$ and $v$ passbands  for two  different
modes ($\ell=0$, $n=1$  and  $\ell=2$, $n=0$) {\it vs} \teff\ 
  for the MS evolution of 1.8~\msun\ models.
Solid circles correspond to complete FST models with the color transformation provided by
 FST atmosphere models. Open-squares refers to complete MLT(\aaa=0.5) with the 
color transformation given by MLT(\aaa=0.5) atmospheres; and crosses corresponds to
complete MLT(\aaa=0.5) and the color transformation given by FST atmosphere models.}

\label{uv-yte}
\end{figure}

\begin{figure}[t]
\resizebox{\hsize}{!}{\includegraphics{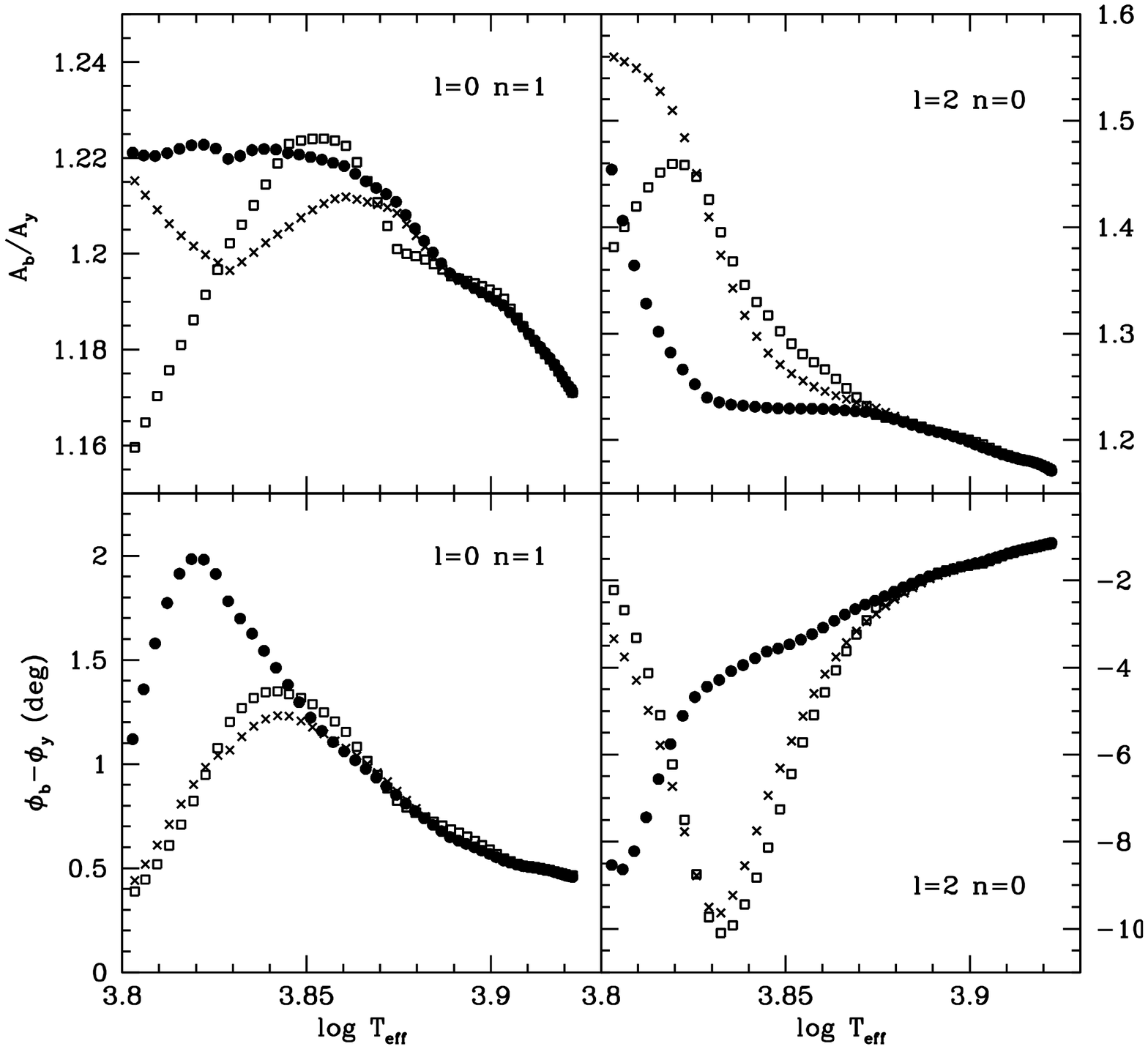}}
\caption{As Fig.~\ref{uv-yte}, but for the $b$ and $y$ passbands.}
\label{by-yte}
\end{figure}

The above  mentioned results directly affect the phase differences 
and amplitude ratios between different photometric passbands. 
More precisely, as  introduced  in Sect.~\ref{sec:colorintro}, two different ingredients allow us to determine
the magnitude variations in different passbands:  the non-adiabatic normalized amplitudes 
and phases of effective temperature variation ($f_{\rm T}$ and $\psi_{\rm T}$); and  the dependence of the monochromatic flux and
of limb-darkening on the effective temperature and gravity. All these quantities, as  mentioned in 
the previous sections,   depend  on the adopted treatment of convection. 

 We consider here three families of atmosphere models: 
 \citet{Kurucz93}'s models  with MLT(\aaa=1.25) treatment, and 
NEMO atmosphere models \citep{Heiter02} with MLT(\aaa=0.5) and FST.

We first recall that, because of the dependence of Eq.~(\ref{mag}) on the degree $\ell$ of the mode, the
comparison between  theoretical and observed photometric amplitude ratios and phase differences makes
$\ell$-identification  possible. On the other hand, different convection treatments lead as well to different results,
so that it is important to check how this affects  mode identification. The classical diagrams for  mode
identification  plot  the amplitude ratio,  $A_{\rm(b-y)}/A_{\rm y}$, versus
 $\phi_{\rm(b-y)}-\phi_{\rm y}$.  
  Fig.~\ref{phi-a} shows the theoretical predictions for  models with $T_{\rm eff}=7400$~K, for modes with 
  degrees $\ell=0$ to $3$ and frequencies close to  the fundamental radial mode.  The results correspond to
different models: one with   MLT($\alpha=0.5$) treatment in the interior and in the atmosphere, a
second one with  
MLT($\alpha=0.5$) in the interior but MLT($\alpha=1.25$) in the atmosphere and a third one with FST treatment
both in the interior and the atmosphere. We see that the regions corresponding to different $\ell$ are still 
clearly disjoint in this diagram when different convection treatments are considered. Moreover, we note
that the typical  precision of photometric methods, i.e. a few degrees for the phases and  
a few percents 
for the amplitude ratios, would allow in principle to discriminate between different convection
treatments. For 
the highest frequencies, however, this discrimination  will be possible only for 
$\ell > 2$ (Fig.~\ref{phi-15}).

A second important point is to examine the respective effects of  
the non-adiabatic calculations and of the atmosphere models on photometric amplitudes and phases.
In Figs.~\ref{uv-yte} and \ref{by-yte} we plot, for three model families, the amplitude ratios and 
the phase differences between two  passbands as a function of the effective temperature. 
Full circles correspond to models where:
the FST treatment of convection  is adopted both in the interior and the atmosphere, 
and  the monochromatic fluxes and limb-darkening, used in the transformation from theoretical to 
observational plane, were computed with the same kind of model atmosphere as in the 
stellar modeling. Open squares correspond to ``complete''  MLT($\alpha=0.5$)  models, and 
as in the previous case, to a consistent transformation to the observational plane. Finally, the crosses 
correspond to a hybrid case where 
we take the non-adiabatic results  $f_{\rm T}$ and $\psi_{\rm T}$ of ``complete''  MLT($\alpha=0.5$)  models,
 but we use the color
transformations (Eq. (\ref{mag})) obtained with the FST atmospheres.

Fig.~\ref{uv-yte} shows  the 
amplitude ratio and phase differences between $u$ and $v$ passbands, and
Fig.~\ref{by-yte} those between $b$ and $y$ passbands.
 Note that scales and ranges in those figures are not the same and that
the effect of different convection treatments are much larger for
the $u$ and $v$ passbands than for the $b$ and $y$ ones.
 We remark in both sets of figures an other interesting result: 
MLT and hybrid models show very close phase differences, 
 and hence, the photometric phase 
differences  mainly provide information about the non-adiabatic results. 
As shown in the previous section,  $\psi_{\rm T}$
essentially depends  on the temperature gradient in the HCZ. Therefore, this gradient can 
be constrained
by comparing the theoretical and observed multi-color photometric phase differences.

\begin{figure}
\resizebox{\hsize}{!}{\includegraphics{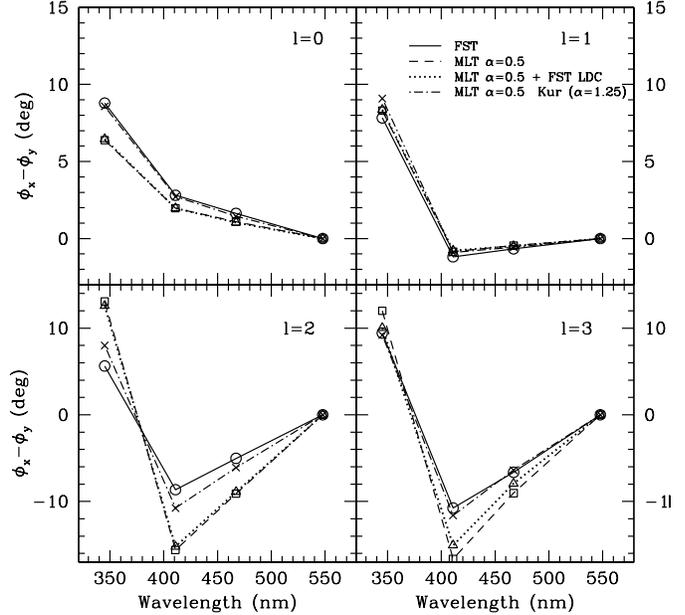}}
\caption{Theoretical monochromatic phase differences $\phi_{u,v,b,y}-\phi_y$ 
as a function of wavelength for 1.8~\msun\ models with \teff=6800~K
and \logg=3.816. Each panel corresponds to a different degree $\ell$ for frequencies close to the fundamental mode.
Solid line: FST interior, FST atmosphere structure and color transformation. Dashed line: MLT(\aaa=0.5) interior, MLT(\aaa=0.5) 
atmosphere structure and color transformation. dotted line: MLT(\aaa=0.5) interior, MLT(\aaa=0.5) atmosphere structure
and  color transformation and limb-darkening coefficients derived from FST atmosphere models (labeled by: FST LDC).
Dot-dashed line:  MLT(\aaa=0.5) interior,  Kurucz MLT(\aaa=1.25)  atmosphere structure and
color transformation.}
\label{phase777}
\end{figure}

The differences between MLT and FST treatments are greatest at  \teff$\sim$6800~K. 
We choose a 1.8~\msun\ star, with $\log$~\teff=3.83 and \logg=3.81, and we compare the  effect of
different external layers treatment on the monochromatic phase differences and amplitude ratios. 
In fig.~\ref{phase777} 
we show  the phase differences for the {\it ubvy} bands as a function of 
the wavelength (a separate panel for each degree  $\ell=0$, 1, 2 and 3). 
Different curves  corresponds to ``complete''  FST and  MLT(\aaa=0.5)
models,  MLT with FST color transformations, and MLT(\aaa=0.5) with 
Kurucz-MLT(\aaa=1.25) atmospheres and the corresponding color transformations.
  
\begin{figure}
\resizebox{\hsize}{!}{\includegraphics{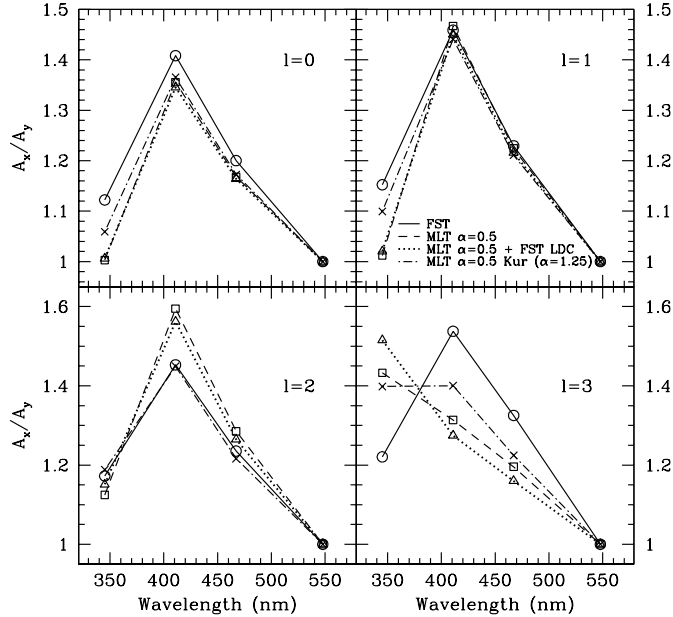}}
\caption{As Fig.~\ref{phase777} but for monochromatic amplitude ratio $A_{u,v,b,y}/A_y$ .}
\label{amplitud777}
\end{figure}

\begin{figure}
\resizebox{\hsize}{!}{\includegraphics{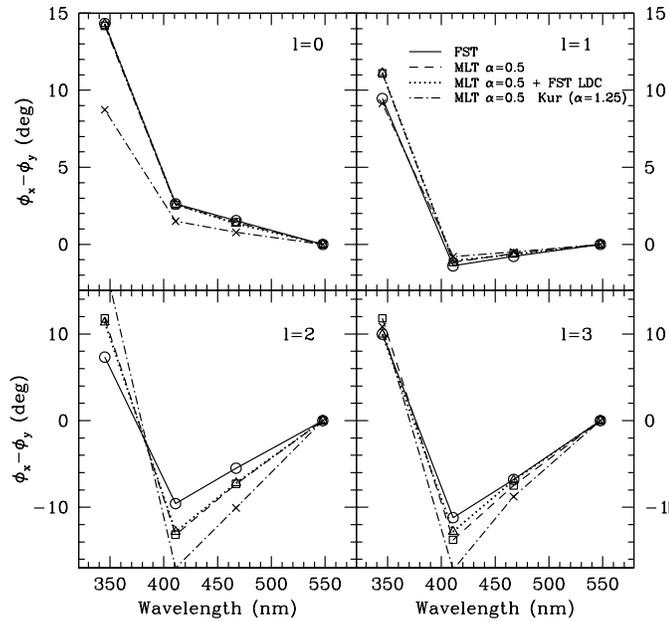}}
\caption{As Fig.~\ref{phase777} but for models 1.8~\msun\ models with \teff=6750~K and \logg=3.802.}
\label{phase780}
\end{figure}

\begin{figure}
\resizebox{\hsize}{!}{\includegraphics{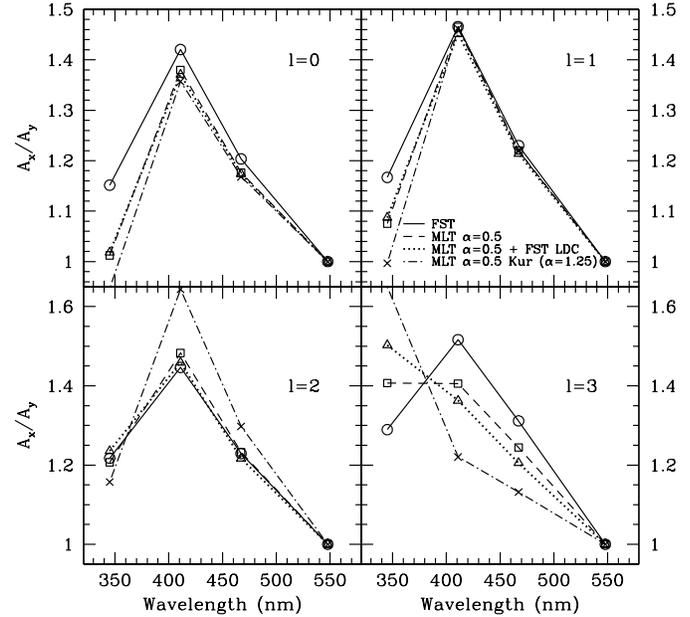}}
\caption{As Fig.~\ref{amplitud777} but for models 1.8~\msun\ models with \teff=6750~K and \logg=3.802.}
\label{amplitud780}
\end{figure}

We confirm that the  atmosphere models used in color transformation do not
have a significant impact on 
the monochromatic phase difference (even if Kurucz-MLT(\aaa=1.25) tables --not included in Fig.~\ref{phase777}--
are used), though the effect increases with $\ell$. 
  The same type of plots of fig.~\ref{phase777} are displayed in fig.~\ref{amplitud777} for the monochromatic 
  amplitude ratio.
 As it  was already shown in Figs.~\ref{ft} and \ref{psit}, the differences in $f_{\rm T}$ and $\Psi_{\rm T}$ between
the different treatments of convection, decrease as mode frequency decrease. That is the reason why 
the ratios $A_{\rm x}/A_{\rm y}$ and phase differences $\phi_{\rm x}-\phi_{\rm y}$ for the modes with frequencies closest 
to the fundamental one presented in figs.~\ref{amplitud777} and \ref{phase777} are so similar for 
the different treatments of the external layers;
and,  why the differences increase with the frequency (or degree $\ell$). 
We think that we must not over-interprete the results obtained for models mixing MLT(\aaa=0.5) in the interior
with Kurucz-MLT(\aaa=1.25) atmospheres. First, because, as already mention in Sect.~\ref{nar}, by mixing 
very different convection treatments, we get unforeseeable results such as those shown in Fig.~\ref{psit}, where the
large value of $\Psi_{\rm T}$ is probably due to a large contribution to the temperature gradient by changing
from \aaa=0.5 to \aaa=1.25 if that happens in an over-adiabatic layer.
Moreover, we must take into consideration the fact that the Kurucz grid is sparser than 
the NEMO one, and the derivatives with respect to \teff\ and \logg\ involved in $\phi_{\rm x}$ and $A_{\rm x}$ are hence
determined with a larger uncertainty. These facts  can be seen in  Figs.~\ref{phase780} and \ref{amplitud780} which
are similar to Figs.~\ref{phase777}, \ref{amplitud777}, but for a 
slightly different  1.8~\msol evolutionary state. These stellar models are only 50~K cooler than the previous ones,
and their gravity is different by $\Delta \log g=0.014$. Now, the behavior of 
$A_{\rm x}/A_{\rm y}$ and $\phi_{\rm x}-\phi_{\rm y}$ 
for the  MLT(\aaa=0.5)+Kurucz-MLT(\aaa=1.25) model has changed, and the differences with FST models have significantly
increased.

 As  the amplitude ratios are concerned,  both the atmosphere models and non-adiabatic results  
 can play a significant role, mainly for higher degree modes.
   Information on the flux and limb-darkening could thus be extracted from 
the amplitude ratios if the  non-adiabatic predictions are known with a sufficiently high degree of confidence. 



\section{Application to the \sct star AN~Lyn}
\label{sec:ANlyn}
\begin{figure}
\resizebox{\hsize}{!}{\includegraphics{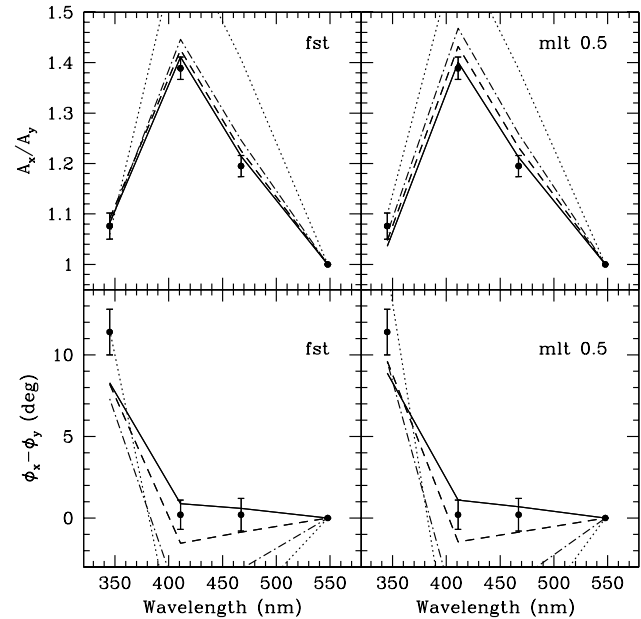}}
\caption{Photometric amplitude ratios $A_{u,v,b,y}/A_y$ (lower panels), and photometric phase
differences $\phi_{u,v,b,y}-\phi_y$ (upper panels) for different models of AN~Lyn.
Lines refers to theoretical predictions for $\ell=0$ (solid line), $\ell=1$ (dashed line),   $\ell=2$ 
(dot-dashed line) and $\ell=3$ (dotted line), and the error bars represent the observations for
$\nu_1$.}
\label{ab-yte}
\end{figure}

AN~Lyn  is a medium amplitude \sct\ star discovered by \citet{Yamasaki81}. Subsequent
investigations  revealed that it is a multi-periodic \sct \citep{Rodriguez97} with a 
peculiar light curve \citep{Rodriguez97b}.
\citet{Rodriguez97} identified three independent frequencies:
 $\nu_1=10.1756$~c/d, $\nu_2=18.1309$~c/d and $\nu_3=9.5598$~c/d, and two of them 
($\nu_1$ and $\nu_2$) have been confirmed by \citet{Zhou02}.
Moreover, the amplitude of the dominant frequency ($\nu_1$) seems to show variations with a
 time-scale of years \citep{Zhou02}.
Precise amplitudes and phases in Str\"omgren $uvby$ photometry were derived by
\citet{Rodriguez97},   we will consider these observational results to 
apply our theoretical models.

The Str\"omgren indices, together with the calibration software TempLoGv2 \citep{StutzNedwich},
provide the following global parameters: \teff=7150~K, \logg=3.65 and [M/H]=-0.17,   and 
taking into account six  different calibrations: \teff=$7100\pm 150$~K,  $\log g=3.5\pm0.35$
and  [M/H]=-0.17$\pm0.03$.
Even if the metallicity resulting from this calibration is sub-solar, we shall do the computations
with classical solar composition, and NEMO atmospheres for [M/H]=0 and microturbulence velocity 
$\xi_{\rm t}=2$~kms$^{-1}$. 
There are several  reasons for these choices:
first of all,  the value of metallicity provided by the photometric calibration is quite uncertain.
Second, \cite{Barban03} computed the limb-darkening coefficients only for that metallicity and $\xi_{\rm t}$; and finally,
the goal of this section is not a precise fitting of the AN~Lyn observations but to analyse how the choice 
of different treatments of the convective outer layers may affect the mode identification.

We have computed a small grid of stellar models for $ Z=0.02$, $X=0.7$ chemical composition, and
masses  from 1.70 to 2.20~\msun. The best fit of \teff~(7166~K), \logg~(3.76),
 main frequency and photometric features 
(amplitude ratios and phases) is obtained for M=2.0~\msol\ at $\tau=1.10^9$~yr, $X_{\rm c}=0.145$, 
a mass of the convective core  $M_{\rm cc}/M=0.18$, and a density contrast $\rho_{\rm c}/\bar{\rho}\sim 860$.
As in previous papers concerning AN~Lyn, we identify  $\nu_1$ as a $n=2$ radial mode. 
 Its  second frequency however is affected by too large errors  to be surely identified.
 For this 2~\msol\ model, this frequency could correspond to a $\ell=3$ $n=2$ mode.

The four panels of Fig.~\ref{ab-yte} show  for the different ($uvby$) bands  the amplitude ratios and phases, 
which  were computed for complete MLT and FST models, and transformed to the observational plane by 
using the limb-darkening coefficients and  integrals for the corresponding model atmosphere.
It have also checked  the effect of combining a kind of atmosphere (MLT) in the stellar model with the
color  and limb-darkening transformations derived from FST model atmosphere.
As it was already pointed out, the main effect is due to the temperature gradient in the external layers, and the 
effect of using FST or MLT based limb-darkening coefficients is negligible.

 We have also computed MLT(\aaa=1.25) models, using consistently the
Kurucz model atmosphere grids and the corresponding limb-darkening coefficients.
The  temperature gradient derived from this models are unable to predict the 
expected amplitude ratios, especially in the $u$~band. 

Concerning FST or MLT (\aaa=0.5), the differences are very small and not observationally disentangled.
This result is not surprising, given the small differences in the theoretical non-adiabatic quantities
shown in Figs.~\ref{ftte2} and ~\ref{psitte2} for a 2~\msol\ star.

\section{Conclusions}
\label{sec:conclu}

Mode identification methods for  \sct\  pulsations  are based on the non-adiabatic
quantities $f_{\rm T}$ and $\psi_{\rm T}$ which are highly sensitive to the characteristics of the 
surface convective zone. The observed phase-lag  between light
variation  and the displacement   originates in two different regions:
 the HeII partial ionization zone, where the $\kappa$ mechanism 
drives  oscillations, and the HI and HeI partial ionization 
region, which coincides with a convective zone (HCZ).
While the contribution from the HeII region mainly depends  on its depth,
and therefore on the evolutionary state, that of HCZ is  very sensitive to
the temperature gradient as well and, therefore, to the convection treatment.
In this paper we have studied  the role of convection in the external layers by comparing low efficiency
convection models: FST and MLT(\aaa=0.5). The novelty of our analysis 
is that, for the first time,
the model atmosphere used in the pulsation analysis were computed with the same treatment of convection as in the internal structure.

This study has shown that:
\begin {itemize}
\item Even if FST and MLT(\aaa=0.5) model atmosphere provide similar stellar spectra and limb-darkening 
\citep{Heiter02,Barban03}, the  non-adiabatic observables $f_{\rm T}$ and $\psi_{\rm T}$ clearly reflect the 
thermal structure of the inner layers of FST and MLT(\aaa=0.5) over-adiabatic regions, showing   differences
 of $\sim 20\degr$ in the phase lag.

\item Differences increase as \teff\ decreases. This is due to the different convective fluxes in 
FST and MLT treatments.

\item For low efficiency of convection, the interaction pulsation-convection is negligible.
So, if convection  efficiency is low enough, 
 we find for the ``frozen convection'' approximation phase-lag values close to results obtained
taking into account the convection-pulsation interaction.
Given that FST is less efficient than MLT, $\psi_{\rm T}$ for FST remains higher than 100$\degr$ for lower \teff\ than
 MLT(\aaa=0.5). 

\item Differences between FST and MLT(\aaa=0.5) also decrease as stellar mass increases.
 With increasing stellar mass, the HCZ becomes more and more shallow and, therefore, convection less and less efficient.

\item The FST results cannot be
reproduced with a single \aaa\ value.
$f_{\rm T}$ and $\psi_{\rm T}$ from FST are bracketed by MLT(\aaa=0.1) and MLT(\aaa=0.3), but 
depending on the frequencies  the FST results are closer to MLT(\aaa=0.1) or  to \aaa=0.3 ones.
Furthermore, for higher \teff\ the FST behavior is closer to that of  MLT(\aaa=0.1), but becomes closer to the MLT(\aaa=0.3) one as \teff\ decreases.

\item The use of Kurucz model atmospheres with \aaa=1.25 in stellar models with \aaa=0.5  have shown that
the differences due to the very external layers can  introduce
an uncertainty in the phase lag of the order of $40\degr$ when compared  with ``complete'' MLT(\aaa=0.5) 
models.

\item As for $\psi_{\rm T}$, photometric phase differences in the Str\"omgren system are mainly sensitive to the
temperature gradient in the HCZ, and are only slighly affected by the color transformation tables and
limb-darkening functions provided by different model atmosphere.
The amplitude ratios, on the contrary, are affected by both the thermal structure and the color transformations.
The $u$ and $v$ bands are in general the most affected: differences in $\phi_u-\phi_y$ of the order of 3-4$\degr$ may be expected
for $\ell$=0 or 2; and differences of 10\% to 20\%  for the amplitude ratio values.

\item Differences between FST and MLT(\aaa=0.5) monochromatic phase differences and amplitude ratios
are also frequency and $\ell$ dependent. For frequencies close to the fundamental mode, 
the sensitivity to the  convection  treatment does not prevent 
the identification of the mode degree $\ell$ from the classical amplitude ratio {\it vs.} phase differences diagram.
But as frequency increases,  $\ell=0$, 1 or 2 mode identification becomes difficult due to the uncertainty 
on FST/MLT convection treatment.
Adding radial velocity data could help, nevertheless, on discriminating between FST and MLT(\aaa=0.5) approaches.

\item The application of our study to the \sct star AN~Lyn provided results in good agreement with previous
publications \citep{Rodriguez97b}. Moreover, given the high mass derived for this star ($\sim 2.0$~\msun) the effect of
convection treatment is small, and it is not possible to discriminate between FST and MLT(\aaa=0.5) on the base of the available observations.

\end {itemize}

In summary, the theoretical photometric amplitude ratios and phase differences between different photometric
passbands are directly related to the non-adiabatic results (see Eq.~\ref{mag}). 
Moreover, different modes probe different layers of the star. Hence, the temperature gradient in the 
superficial convection zone of intermediate mass stars can be constrained by comparing these amplitude 
ratios and phase differences with observations. Such information cannot be obtained from the spectrum alone because
spectral lines probe only region of small optical depth.

\begin{acknowledgements}
We thank  Dr. Rafael Garrido for kindly send  us  his color-transformation program.
J.M acknowledges financial support from the Prodex-ESA Contract Prodex 8 COROT (C90199).
\end{acknowledgements}

\bibliographystyle{aa}
\small
\bibliography{mad}
\end{document}